\documentclass[useAMS,usenatbib,usegraphicx]{mn2e}
\newcommand{\msunh}{h^{-1}{\rm M}_{\solar}}
\newcommand{\solar}{\ifmmode_{\mathord\odot}\;\else$_{\mathord\odot}\;$\fi}

\def\ltsima{$\; \buildrel < \over \sim \;$}
\def\lsim{\lower.5ex\hbox{\ltsima}}
\def\Msunh{\mbox{$h^{-1}$M$_\odot$}}
\def\mpch{\mbox{$h^{-1}$Mpc}}

\def\deg{\ifmmode{^\circ}\else{$^\circ$}\fi}
\def\hGpc{\ifmmode{h^{-1}{\rm Gpc}}\else{$h^{-1}{\rm Gpc}$}\fi}
\def\hkpc{\ifmmode{h^{-1}{\rm kpc}}\else{$h^{-1}{\rm kpc}$}\fi}
\def\hMpc{\ifmmode{h^{-1}{\rm Mpc}}\else{$h^{-1}{\rm Mpc}$}\fi}
\def\hMsun{\ifmmode{h^{-1}M_\odot}\else{$h^{-1}M_\odot$}\fi}

\def\muK{\ifmmode{\mu{\rm{K}}}\else{$\mu$K}\fi}
\def\mum{\ifmmode{\mu{\rm{m}}}\else{$\mu$m}\fi}

\newcommand{\aj}{{AJ}}
\newcommand{\apj}{{ApJ}}

\newcommand{\apjs}{{ApJS}}
\newcommand{\mnras}{{MNRAS}}

\title{On an Analytical Framework for Voids: Their abundances, density profiles and local mass functions}

\author[Patiri et al.]
{S. G. Patiri$^{1}$\thanks{E-mail:
spatiri@iac.es},
J. Betancort-Rijo$^{1,2}$,
F. Prada$^{3}$, 
\\
$^1$
Instituto de Astrofisica de Canarias,
C/ Via Lactea s/n, Tenerife, E38200, Spain
\\
$^2$
Facultad de Fisica, Universidad de La Laguna,
Astrofisico Francisco Sanchez, s/n, La Laguna
Tenerife, E38200, Spain
\\
$^3$
Ramon y Cajal Fellow, Instituto de Astrofisica de Andalucia (CSIC), E-18008, Granada, Spain
\\
}

\begin{document}

\maketitle

\begin{abstract}
We present a general analytical procedure for computing the number density of voids
with radius above a given value 
within the context of gravitational formation of the large scale structure of 
the universe out of Gaussian initial conditions. To this end we develop an 
accurate (under generally satisfied conditions) extension of the
unconditional mass function to constrained environments, which allows us both to 
obtain the number density of collapsed objects of certain mass at any distance 
from the center of the void, and to derive the number density of voids defined 
by collapsed objects. We have made detailed calculations for the spherically
averaged mass density and halo number density profiles for particular voids. 
We also present a formal expression for the number density of voids defined by 
galaxies of a given type and luminosity. This expression contains the probability for a collapsed 
object of certain mass to host a galaxy of that type and luminosity (i.e. the conditional
luminosity function) as a function of the environmental density. We propose a 
procedure to infer this function, which may provide useful clues as to the 
galaxy formation process, from the observed void densities. 

\end{abstract}


\section{Introduction}

It is well known that the distribution of galaxies in the Universe is not uniform.
The galaxies are distributed in filaments and clusters, leaving large regions devoid of
bright galaxies. These regions are known as voids.

The first giant void was the so-called Bo\"otes void found by \citet{Presi}. Subsequently, 
thanks to large redshift surveys, a large amount of these regions were found and analyzed
(e.g. de Lapparent et al. (1986) and Vogeley et al. (1994) in the CfA redshift survey; 
El-Ad et al. (1997) in the IRAS survey; M\"uller et al. (2000) in Las Campanas Redshift Survey;
Plionis \& Basilakos (2002) in the $PSC_{z}$; Croton et al. (2004) and Hoyle \& Vogeley (2004) in the
2dFRGS; Conroy et al.(2005) in the DEEP2 redshift survey)

Many of the theoretical works in the literature about voids  
are based on cosmological N-Body simulations. The first simulations of the dark matter 
distribution (Davis et al. 1985) qualitatively showed the existence of large low density regions, 
but detailed studies of these regions with better resolution are just 
becoming available (Van de Weygaert \& Van Kampen 1993; Gottl\"ober et al. 2003; 
Colberg et al. 2004 using N-Body simulations and Mathis \& White 2002; 
Benson et al. 2003 using semi-analytical models). Aside from the simulations, there are many analytical works dealing 
with underdensities in the mass field (see e.g. Sheth \& van de Weygaert 2004, and references therein) that give
 good descriptions of voids. 
On the other hand, analytical works that define the voids by observable objects (i.e. in point distributions) 
are not common in the literature (White 1979; Otto et al. 1986; Betancort-Rijo 1990). 

An important point about voids is the study of their contents. Despite
the word, voids, of course, are not empty. The first detections of galaxies inside voids
were spirals near the 'border' of previously defined voids like the Bo\"otes (Dey et al. 1990; Szomoru et al. 1996a and 1996b). 
However, extrapolating the morphology-density relation (Dressler 1980) one might expect a population of dwarf
galaxies well inside the voids. Even though such galaxies have not been observed yet, they
will provide, along with the voids statistics, a strong test for the galaxy 
formation models (Peebles 2001). There is some ongoing progress in the studies of galaxy
populations in voids, thanks mainly to the contribution of recent large redshift 
surveys (see e.g. Rojas et al. 2004; Patiri et al. 2005b).

There are strong discussions about what exactly constitute a void and therefore how to
define them. In general, authors define what is a void depending on the studies
they are carrying on. In some works, voids are defined as underdensities in a 
continuous underlying field (Van de Weygaert \& Van Kampen 1993; Aikio \& Maehoenen 1998; Friedmann \& Piran 2001; 
Sheth \& Van de Weygaer 2004; Colberg et al. 2004). In other works, voids are
irregular regions delimited by some kind of galaxies, the so-called 'wall' galaxies 
(e.g. El-Ad \& Piran 1997; Hoyle \& Vogeley 2002; Benson et al. 2003; Hoyle \& Vogeley 2004). 
Even though, these definitions give a very good idea, for example, about the shape of the galaxy distribution, they do not 
provide a particularly powerful tool for statistical inference. For this purpose we need a definition
which does not smears the information contained in the actual object distribution. Note that if we 
filter this distribution in certain scale so as to obtain a continuous field and use it to define
voids, or classify the objects not by an intrinsic criteria but by a distribution dependent
one (e.g. 'wall galaxies'), information is smeared and the ability to discriminate between models by comparing observations
with predictions is diminished. This effect is similar to what happens in regard to binned data: the best test using binned data is 
never better and usually worst than the best test using the raw data.

We define voids as maximal non-overlapping spheres empty of objects with mass above a given one. 
For example, we could define voids as maximal spheres empty of Milky Way-size halos, so that even though, the voids are 
empty of these halos, we can have dwarf halos inside the voids (see Fig.(1)). Otto et al. (1986) and Gottl\"ober et al. (2003) 
also use this kind of definition.

We will focus our work mainly on rare voids (e.g. the giants ones) because the mean number of these voids 
is a very sensitive function of the clustering properties of the objects
that define those regions. This implies that the available statistics on 
voids along with more general statistics like the counts in cell moments may be 
used to obtain accurate information about those clustering properties, 
providing a powerful tool to discriminate between different large scale structure formation models.
To this end the following elements are required: 
first, a handy framework to compute, for given clustering properties, the mean number of voids 
defined by certain kind of objects as a function of radius and,
second,	a precise characterization of the clustering properties which is both 
easy to use in the framework and physically meaningful.

\begin{figure}
\includegraphics[width=\columnwidth]{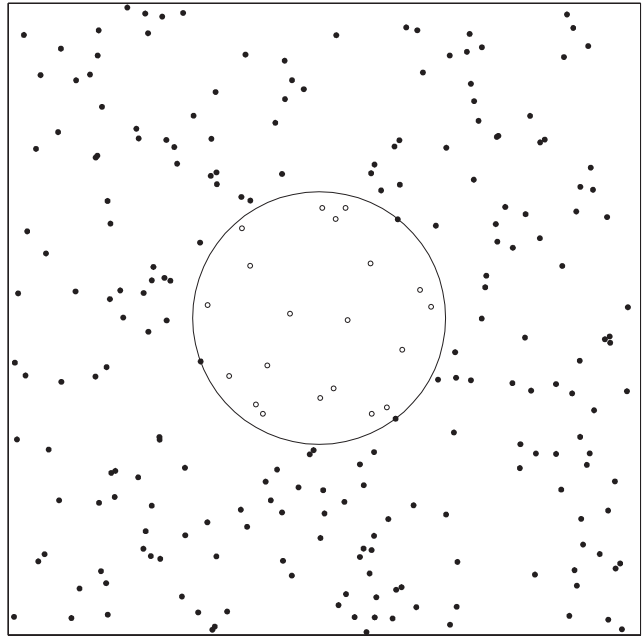}
\caption{Our definition of voids: maximal non-overlapping spheres empty of objects with 
mass above $M$ (filled circles). As it is shown, voids can contain objects with masses 
smaller than $M$ (open circles). Note that we do not show for clarity the smaller object 
outside the void.}
\label{fig:fig05}
\end{figure}

The aim of this work is to provide these elements and assess their efficiency.
We will show how they may be used to infer properties of the processes whereby halos become
galaxies of certain kind from the statistics of voids defined by galaxies of
that kind. To this end we need to express the void probabilities in
terms of the galaxy clustering properties. This can be done in different ways. 
For example, using all galaxy  correlation functions to characterize
their clustering properties we could, in principle, obtain the corresponding void
statistics \citep{White79}. However, in practice, this procedure is not feasible. Furthermore,
even if it were, the information obtained about the clustering properties in
this representation do not have a direct connection with galaxy formation processes.
In the procedure we present in this work, we first compute the probabilities of
voids defined by halos with masses above a given value. This can be done analytically by
combining statistics with purely gravitational dynamics. Then, we describe the clustering 
properties of the galaxies by their relative biasing with respect to halos of the same
mass, which may be expressed in terms of the conditional luminosity function, and obtain an 
expression for the mean number of voids defined by galaxies. For this relative biasing we 
may either use the existing semi-analytic models (for example, \cite{VanDen})
using our formalism in a predictive way, or use this formalism in an inductive
way to determine that biasing from the observations. This will be presented in a formal way in the discussion,
leaving its applications for a future work.

The structure of the work is as follows. In section 2 we use an existing framework 
\citep{Beta90} that allows us to derive the number density\footnote{Technically, the correct term is the {\emph probability density}, which 
is well defined even when it changes substantially over the local mean distance between voids (or relevant objects). 
However, to avoid confusion we shall use instead the term: "number density"} of
voids of a given radius from the
probability that a randomly placed sphere of that radius be empty of the objects
defining the void (the VPF). We also show in section 2 how this probability may be
obtained by means of an expression containing the biasing of halos with respect to mass.
In section 3 we present an extension of the unconditional mass function (UMF) 
\citep{ST99,Jenkins} to constrained environments, which allow us to obtain the 
conditional mass function (CMF) which, in turn, is used to quantify the biasing of halos with respect to mass. 
In section 4 we show, by combining spherical collapse with the CMF,  
how to obtain the mean density profile both for the mass and for the halos within and around voids.
In section 5 we apply our formalism to different cases to obtain void probabilities and their mass density and
halo number density profiles and compare the results with those found in existing numerical simulations.
Finally, in section 6 we discuss how to use our formalism to obtain the probabilities of voids defined
by galaxies.

\section{Probabilities of voids: The framework}
\label{ref:Sec1}

The general framework for evaluating the mean number densities (i.e. probability 
densities) of structures like voids \citep{Preskill,otto} has been applied to the 
standard large scale structure models (i.e. Gaussian initial density fluctuation 
growing gravitationally). The framework we use here is essentially an updated 
and extended version of the one developed by \citet{Beta90}. 
There, the number density, $\bar{P}_{0}(r)$ of non-overlapping empty (of the objects defining the voids)
spheres with radius $r$ is given by:

\begin{equation}
\bar{P}_{0}(r)=\frac{3\pi^2}{32}\frac{(\bar{n}'V)^3}{V}P_{0}(r) (1+\mathcal{O}((\bar{n}'V)^{-1}))  \label{eq:eq1}
\end{equation}
with

\begin{equation}
\bar{n}'=\frac{1}{4\pi r^2}\frac{d\ln P_{0}(r)}{dr} ;\qquad V=\frac{4}{3}\pi r^3    \nonumber
\end{equation}
and

\begin{equation}
P_0(r)=\int_{-1}^{\infty} e^{-(\bar{n}V)(1+\delta _{N}(\delta))}~P(\delta/r)~d\delta      \label{eq:eq2} 
\end{equation}
where $P_{0}(r)$ is the probability that a randomly placed sphere of radius $r$ be empty (this is the
so-called Void Probability Function, VPF), $\bar{n}$ is the mean number density of the objects defining
the void, $\delta _{N}(\delta)$ is the fractional fluctuation of the 
mean number density of these objects within a sphere of radius $r$ as a function of the actual fractional 
mass density fluctuation within that sphere, $\delta$, and $P(\delta/r)$ is the probability distribution 
for the values of $\delta$ within a randomly chosen sphere of radius $r$. The bias of halos with 
respect to mass is contained within $\delta_{N}(\delta)$; in the non-biased case (which corresponds to the low 
halo mass limit) we have simply $\delta_{N}$ equal to $\delta$.

Equation (\ref{eq:eq1}) (without the last parenthesis) is valid for rare events, that is, when the mean distance between
voids is much larger than the radius, which imply:

\begin{equation}
k\equiv\bar{n}'V  >> 1 \nonumber
\end{equation}
               
In fact, when $k$ is larger than about $3.5$, equation (\ref{eq:eq1}) with only the zeroth order term in the last 
parenthesis is sufficiently accurate. Extending eq.(\ref{eq:eq1}) to smaller values of $k$ (i.e. obtaining the 
terms of order $k^{-1}$) is straightforward (but complex), however, eq.(\ref{eq:eq1}) (without the last parenthesis) 
will be enough to our purposes because the most relevant constraints for galaxy formation models comes from not too common
voids (i.e. $k$ sufficiently large).

The exponential in eq.(\ref{eq:eq2}) represents the probability that a randomly placed sphere of radius $r$ be
empty of the relevant objects, when the mean fractional density fluctuation within the sphere
take the value $\delta$. Multiplying this quantity by the probability for a randomly placed sphere of radius $r$
to have an inner mean fluctuation between $\delta$ and $\delta + d\delta$, $P(\delta/r)~d\delta$, and integrating
over all possible values of $\delta$ we obtain $P_{0}(r)$.

It must be noted that the exponential in Eq.(\ref{eq:eq2}) gives correctly the probability that the sphere be empty only
when the clustering of the relevant objects conforms to a non-uniform random Poissonian process (Peebles 1980).
This is the case to a very high accuracy both when the objects are mass particles or halos of a given mass, on scales 
(as those of voids) which are much larger than the size of the halos. For galaxies, the model might not be so good.
In section 6 we show how to modify expression (\ref{eq:eq2}) to be valid for objects whose clustering properties conform
to any possible interesting model.

For mass particles $\delta_{N}=\delta$, so that eq.(\ref{eq:eq2}) is particularly simple. From this expression it is easy to see 
that as the density of particles ($\bar{n}$) increases, the size of the voids goes to zero. 

The probability distribution, $P(\delta/r)$, for a given power spectra is given, for any value of $\delta$
in \citet{Beta02}. However, since the evolution of large voids is well described by the spherical model, 
we may use in eq.(\ref{eq:eq2}) the spherical approximation to $P(\delta/r)$, also given in that reference 
with an error that for the voids under consideration is not very relevant, although in some cases, when high accuracy is 
required, the full $P(\delta/r)$ may be needed. So, we shall use for $P(\delta/r)$:

\begin{eqnarray}
P(\delta/r)d\delta \equiv P(\delta_{l}/r)d\delta_{l} = \nonumber  \\ \frac{1}{\sqrt{2\pi}}\frac{\exp \left(-\frac{1}{2}\frac{\delta 
_l^2}{\sigma^2(r(1+\delta)^{1/3}))}\right)}{(1-(1-\frac{\alpha}{2})(1-(1+\delta)^{-1/3}))^{-3}}d\bigg( \frac{\delta 
_l}{\sigma(r(1+\delta)^{1/3}))}\bigg)       \nonumber      \label {eq:eq3}
\\
\end{eqnarray}

\begin{equation}
\alpha(r)=-\frac{1}{3}\frac{d\ln\sigma^2(r)}{d\ln r} \simeq 0.54+0.173\ln\bigg(\frac{r}{10h^{-1}Mpc}\bigg)  \nonumber
\end{equation}
where $\sigma^2(r)$ is the variance of the linear density fluctuation with a top-hat 
filtering on a scale $r$ as a function of $r$; $\delta _l$ is the linear value of the 
density fluctuations which is related to $\delta$ by the spherical model. 
For $\Omega _m=0.3$ and $\Omega _\lambda=0.7$ we have:

\begin{eqnarray}
\delta = D(\delta _l) \equiv 0.993[(1-0.607(\delta _{l}-6.5\times 10^{-3} \times \nonumber \\ 
\times (1-\theta(\delta_{l})+\theta(\delta_{l}-1.55))\delta_l^2))^{-1.66}-1]\nonumber  \label{eq:eq4}
\\
\end{eqnarray}
where

\begin{equation}
 \theta(x)=\left \{ \begin{array}{ll}
  1 & \rm{if~x>0}\\
  0 & \rm{if~x\leq 0}
\nonumber \end{array} \right.   
\end{equation}

Using eq.(\ref{eq:eq3}) and eq.(\ref{eq:eq4}) in eq.(\ref{eq:eq2}) and changing the integration variable to
$\delta _l$, we have for $P_0(r)$:

\begin{equation}
P_0(r)=\int_{-\infty}^{\infty} e^{-\bar{n}V~(1+\delta _{N}(\delta _{l}))}~P(\delta _{l}/r)~d\delta _l  \label{eq:eq5}
\end{equation}
where $\delta_{N}(\delta _l)$ is the mean fractional fluctuation within $r$ of the number
density of the objects defining the void as a function of the linear fractional density fluctuation within $r$. 
Alternatively, integrating over $\delta$ we may write:

\begin{equation}
P_0(r)=\int_{-1}^{\infty} e^{-\bar{n}V~(1+\delta _{N}(DL(\delta)))}~P(\delta)~d\delta \label{eq:eq6b}
\end{equation}
where

\begin{eqnarray}
\delta_{l} \equiv DL(\delta) = \frac{\delta_{\rm c}}{1.68647}\bigg[ 1.68647 - \frac{1.35}{(1+\delta)^{2/3}} - 
\frac{1.12431}{(1+\delta)^{1/2}} + \nonumber \\ + \frac{0.78785}{(1+\delta)^{0.58661}}\bigg]   \label{eq:eq6}
\end{eqnarray}

This expression for $DL(\delta)$ \citep{ST02} corresponds to the same cosmological
parameters as eq.(\ref{eq:eq4}) with $\delta_{\rm c}=1.676$ and is the inverse function of eq.(\ref{eq:eq4}).
Note that we write $\delta _{N}(\delta _{l}(\delta))$ rather than simply $\delta_{N}(\delta)$ because it 
is $\delta_{N}(\delta_l)$ that we may compute directly (see next section), while $\delta_{N}(\delta)$ is 
obtained through the dependence of $\delta_l$ on $\delta$. Here we shall use the first expression (Eq.(\ref{eq:eq5})) 
for $P_{0}(r)$ while the second must be used when the exact $P(\delta)$, rather than the spherical 
collapse approximation, is needed.
Using eq.(\ref{eq:eq5}) (or eq.(\ref{eq:eq6b})) in eq.(\ref{eq:eq1}) we obtain the mean density of
non-overlapping empty sphere of radius $r$, $\bar{P}_{0}(r)$. The number density of voids so
that the largest sphere that they can accommodate have radii between $r$ and $r+dr$,
$n(r)dr$, is related to $\bar{P}_{0}(r)$ by:

\begin{eqnarray}
\bar{P}_{0}(r) = n(\geq r)+n(\geq 2 \alpha_1 r)+n(\geq(1+\frac{2}{\sqrt{3}})\alpha_2r)+ \nonumber \\ +
n(\geq(1+\sqrt{3/2})\alpha_3 r)+n(\geq(1+2\sqrt{2/3})\alpha_4 r)+ \nonumber \\ +
6n(\geq 2.738\alpha_5 r)+\cdots \nonumber \label{eq:eq7}
\end{eqnarray}  

\begin{equation}
n(>r)\equiv\int_{r}^{\infty} n(r)~dr    \nonumber
\end{equation}  
where $\alpha_i$  depends on the mean ellipticity of the voids and may be taken equal to $1$ 
without losing much accuracy.     
But, in fact, in the interesting cases ($k >> 1$), $n(2r)/n(r) << 1$ , so we
may write:

\begin{equation}
\bar{P}_{0}(r) \simeq n(\geq r)  \quad;\quad n(r) \simeq - \frac{d}{dr} \bar{P}_{0}(r)   \label{eq:eq8}
\end{equation}  	     

The mean number, $N(r,\bar{V})$, of voids
with radius larger than $r$ within the volume of observation $\bar{V}$ is then: 

\begin{equation}
N(r,\bar{V})=\bar{V}~n(>r)\cong~\bar{V} \bar{P}_{0}(r)   \nonumber
\end{equation}

\section{Derivation of $\delta_{N}(\delta_{l})$}
\label{ref:Sec2}

\subsection{Steps to follow}

To derive the mean fractional number density fluctuation within a sphere, $\delta_{N}$,
as a function of the mean linear fractional density fluctuation within it, $\delta_{l}$,
we first obtain the mean fractional fluctuation of the number density of proto-halos 
within the sphere before the sphere expands in comoving coordinates, $\delta_{ns}$, as a 
function of $\delta_{l}$. $\delta_{ns}$ may be called the statistical fluctuation, since it
is due to the clustering of the protohalos in the initial conditions before they move with mass.
To obtain  $\delta_{ns}(\delta_{l})$ we need a framework that enables us to obtain the 
conditional mass function (CMF) of collapsed objects $n_{c}(m,Q,q,\delta_{l})$ at a distance $q$ from 
a point such that the mean density within a sphere of radius $Q$ (radius of the void) centered 
at this point is $\delta_{l}$ (the linear value; $\delta=D(\delta_{l})$ is the actual one). $Q$ and 
$q$ denote the Lagrangian radius; for Eulerian radius (i.e. present comoving radius) we use 
respectively $R$ and $r$. We then have for $\delta_{ns}$:

\begin{equation}
1+\delta_{ns}(m,Q,\delta_l)=\frac{1}{N_{u}(m)}\bigg[\frac{3}{Q^3}\int_{0}^{Q}N_{c}(m,Q,q,\delta_{l})~q^2~dq\bigg]  
\label{eq:eq9}
\end{equation} 
	
\begin{equation}
N_{u}(m)\equiv\int_{m}^{\infty}n_{u}(m)~dm   \nonumber
\end{equation}

\begin{equation}
N_{c}(m,Q,q,\delta_{l})=\int_{m}^{\infty}n_{c}(m,Q,q,\delta_{l})~dm  \nonumber
\end{equation}	
where $n_{u}(m)$ is the unconditional number density of collapsed object with mass
$m$. The bracketed expression is the mean value within the sphere of radius $Q$ of the number
density, $N_{c}$, of collapsed objects with masses above $m$ when the mean linear fractional 
mass density fluctuation within it is $\delta_l$.

As indicated in eq.(\ref{eq:eq9}), $\delta_{ns}$ depends, in principle, on $m$, $Q$ and $\delta_{l}$. However, it
may be shown on general grounds (and we have fully checked) that $n_{c}$, $N_{c}$ depend on $q$
and $Q$ almost entirely through $q/Q$. The reason for this lays in on the goodness of the approximation 
given in expression (\ref{eq:eq110}). As mentioned bellow this expression, there is a small residual 
dependence on $Q$, but it is completely negligible within the relevant range of $Q$ values (less than a 
factor 2). Then it follows from eq.(\ref{eq:eq9}) that $\delta_{ns}$ is independent of $Q$, since, using
the change of variable $u=q/Q$, we may write this equation in the form:

\begin{equation}
1+\delta_{ns}(m,\delta_l)=\frac{1}{N_{u}(m)}\bigg[3 \int_{0}^{1}N_{c}(m,u,\delta_{l})~u^2~du\bigg]  
\label{eq:eq999}
\end{equation} 

To obtain the fluctuation in Eulerian coordinates, $\delta_{N}$, we only need to note that as the void expands the 
initial fractional halo number density further diminishes by the factor $(1+\delta)$:

\begin{equation}
1+\delta_{N}(m,\delta_{l})=(1+\delta_{ns}(m,\delta_{l}))(1+\delta(\delta_{l})) \label{eq:eq122}
\end{equation}  

This is the expression that we use in eq.(\ref{eq:eq5}) to obtain $P_{0}(r)$.
The unconditional mass function of collapsed objects. $n_{u}(m)$ is given with high accuracy 
by the unconditional Sheth \& Tormen approximation (Sheth \& Tormen 1999, 2002; hereafter ST):

\begin{eqnarray}
    n_{\rm uST}(m,\delta_{c},\sigma(m)) &=& - \left( \frac{2}{\pi} \right)^{1/2}
    A \left[ 1 + \left( \frac{a \delta_{\rm c}^2}{\sigma^2} \right)^{-p}
    \right]  \label{eq:eq10} \\
    & \times& a^{1/2} \frac{\varrho_{\rm b}}{m}
    \frac{\delta_{\rm c}}{\sigma^2} \frac{{\rm d} \sigma}{{\rm d} m}
    \exp \left( - \frac{a \delta_{\rm c}^2}{2 \sigma^2} \right)   \nonumber
\end{eqnarray}
where $A=0.322$, $p=0.3$ and $a=0.707$, $\varrho_b$ stands for the background density,
$\sigma$ is the $rms$ linear mass density fluctuation and $\delta_c$ is the value of 
$\delta_l$ corresponding to collapse in the spherical model which for the cosmological 
parameters that we use ($\Omega_m$=0.3, $\Omega_\lambda$=0.7) is 1.676.
Our problem now is to obtain a similarly accurate approximation to the
conditional mass function, $n_{c}(m,q,Q,\delta_{l})$, so that we can use it in Eq.(\ref{eq:eq999}).

\subsection{Local constrained mass function}

 \subsubsection{Why do we need a new approach to the CMF ?}

Expression (\ref{eq:eq999}) gives the Lagrangian constrained accumulated mass function, $N_{cL}(m,\delta_{l})$, 
averaged within a sphere with mean inner linear underdensity $\delta_{l}$ and radius $Q$ (whose dependence on $Q$ has been neglected)
given the local Lagrangian accumulated mass function at a distance $q$ from the center of that sphere, $N_{c}(m,u,\delta_{l})$:

\begin{equation}
N_{cL}(m,\delta_{l})= N_{u}(m)(1+\delta_{ns}(m,\delta_{l}))  \nonumber
\end{equation}

Thus, the accumulated Eulerian mass function averaged within the sphere, $N_{cE}$, which is the ordinary
mass function, is given by:

\begin{equation}
N_{c}(m,\delta_{l}) \equiv N_{cE}(m,\delta_{l})= N_{u}(m)(1+\delta_{ns}(m,\delta_{l}))(1+\delta(\delta_{l}))  \nonumber
\end{equation}

These expressions involve an integral within the sphere of the local Lagrangian constrained mass function,
$n_{c}(m,q,Q,\delta_{l})$. Thus, this last function is necessary to obtain both $N_{c}(m,\delta_{l})$ and 
through expression (\ref{eq:eq122}), $\delta_{N}(m,\delta_{l})$.
There are several approaches (Sheth \& Tormen 2002; Gottl\"ober et al. 2003; Golberg et al. 2004) giving 
rough approximations and providing reasonable fitting formula for $N_{c}(m,\delta_{l})$, however none of them 
provides directly (without fitting) an expression sufficiently accurate to allow us to evaluate the 
void number densities, which depends very sensitively on $\delta_{N}(m,\delta_{l})$. This is due to 
the fact that the local mass function changes substantially from the center of the sphere ($q=0$) to 
its boundary ($q=Q$). In fact, for any $\delta_{l}$, the mean value of $n_{c}(m,q,Q,\delta_{l})/n_{u}(m)$ 
within the sphere is approximately the square root of its value at the center and the value of this quantity at the 
boundary is almost the cubic root of its value at the center. 
So, computing the mass function at the center instead of the mean value, or assuming that the interior
of the sphere may be replaced by a homogeneous environment characterized by the mean properties of the
actual one do not lead to sufficiently good results.

There have been several attempts to derive analytically the CMF. Arguably, the most motivated one is that combining
the excursion set formalism (Appel \& Jones 1990; Bond et al. 1991) with ellipsoidal dynamics (Sheth \& Tormen 2002). 
However, this procedure is not appropriate to our purposes because, by construction, it gives the CMF at the center of
the sphere, which, as we stated before, is quite different from the mean within the sphere, which is the one
that we need. In principle, one could repeat the ST derivation at $q=0$ for any value of $q$ and average over 
the sphere as indicated in Eq.(9), but this implies a rather complex problem that can not be solved
without some approximations. Furthermore, even if the problem could be treated exactly it would provide
at most a good fitting formula where some parameters have to be slightly modified (with respect to
those given directly by the formalism) to match numerical simulations, as, indeed, have already been done
for the unconstrained case (Sheth \& Tormen 2002).

In another approach (Gottl\"ober et al. 2003), the interior of the sphere is treated as a homogeneous
environment and the unconditional mass function is rescaled to it. But, leaving aside some queries
about the motivation for this procedure, in fact, it disagrees substantially with the simulations
for large masses.

Summarizing, neither the available CMF's nor any other we can envisage derived from simple considerations 
can a priori be expected to give results which are sufficiently accurate to our purposes.
Fortunately, although we can not directly obtain analytically a satisfactory CMF, we may analytically
extend the UMF through a procedure which is, in practice, exact.

\subsubsection{CMF: extending the Unconditional Mass Function}

To obtain the CMF, we simply note that, as long as the local 
evolution at a conditioned point is the same as at an unconditioned one, the conditional local mass function 
of collapsed objects may be derived from the statistical properties of the local linear field 
in the same way as the unconditional one.

As to the validity of this assumption, three reasons may be advanced:

\begin{itemize}

 \item{Given the large difference between the scale of the constraint (that of the void) and the scale corresponding 
to the masses we consider, the conditional shear distribution (of the field filtered on the scale of those masses) 
can not be very different from the unconditional one}

 \item{The profile of the linear density field within the void is not too steep. This means that the mean negative value 
of the shear in the radial direction imposed by this profile is rather small (it would be strictly zero for a flat 
profile). So, the departure of the local shear distribution from the unconditional one is smaller than implied in 
general by the first consideration}

 \item{It must be noted that the shear distribution plays a secondary role with respect to the trace (of the local 
velocity field tensor) in determining the mass distribution of collapsed objects. The difference between ST and the 
PS formalisms is due to the fact that in the former, the shear distribution is taken into account. The 
difference between the results of both formalisms is not that large (less than a factor 2). So, the small change in 
the shear distribution within a void which, according to the two previous considerations, is small, should lead to a 
negligible error for our extended mass function}

\end{itemize}

\begin{figure}
\includegraphics[width=\columnwidth]{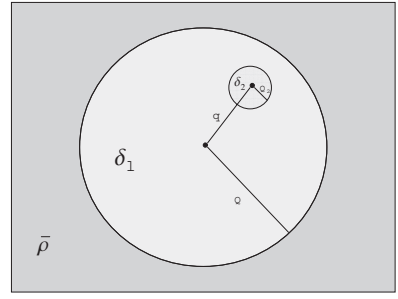}
\caption{In order to compute the CMF evaluated at certain mass at 
a distance $q$ from the center of a sphere of radius $Q$ with mean inner linear underdensity $\delta_{1}$, we 
consider the probability distribution at $q$ for the mean linear underdensity ($\delta_{2}$) within a sphere 
with radius $Q_{2}$ corresponding to the mass under consideration (see text for details).}
\label{fig:fig1}
\end{figure}

That is, if the constrained field behaves ``locally" as an isotropic uniform random Gaussian field
(with locally defined mean and power spectra), or, alternatively, if the shear distribution
of the linear velocity field is at a constrained point equal to that at a randomly chosen one, we
may obtain the local mass function using the Unconstrained Mass Function (UMF) for this local field. 
The relevant statistical property is the probability distribution, $P(\delta_{2}/\delta_{l},q,Q)$, 
for the linear density fluctuation, $\delta_{2}$, on scale $Q_{2}$ (that of the halos considered) 
at a distance $q$ from the center of a sphere of radius $Q$ (the protovoid) with mean inner linear 
density fluctuation $\delta_l$ (see the conceptual diagram in figure (\ref{fig:fig1})).

For a Gaussian field we have:

\begin{equation}
P(\delta_{2}/\delta,q,Q)=\frac{ \exp \left( -\frac{1}{2} \frac{\big(\delta_{2}-\delta_{l}\frac{\sigma_{12}}{\sigma_{1}^{2}} 
\big)^2}{\big(\sigma_{2}^{2}-\frac{\sigma_{12}^{2}}{\sigma_{1}^{2}}\big)}
\right)}{\sqrt{2\pi}\big(\sigma_{2}^{2}-\frac{\sigma_{12}^{2}}{\sigma_{1}^{2}}\big)^{1/2}}  \nonumber
\end{equation}
where

\begin{equation}
\sigma_{1}^{2}\equiv<\delta_{1}^{2}>=\sigma^{2}(Q)   ;\qquad \sigma_{2}^{2}\equiv<\delta_{2}^{2}>=\sigma^{2}(Q_{2})  \nonumber
\end{equation}

\begin{equation}
\sigma^{2}(x)=\frac{1}{2\pi^{2}}\int_{0}^{\infty}\mid\delta_{k}\mid^{2}~W^2(xk)~k^{2}~dk  \nonumber
\end{equation}

\begin{eqnarray}
\sigma_{12}\equiv\sigma_{12}(q,Q,Q_{2})= \\ \frac{1}{2\pi^{2}q}\int_{0}^{\infty}\mid\delta_{k}\mid^{2}~W(Qk)~W(Q_{\rm 2}k)~sin(kq)~k~dk   \nonumber
\end{eqnarray}

\begin{equation}
W(x)=\frac{3}{x^{3}}(sin~x~-x~cos~x~)   \nonumber
\end{equation}
where $\mid\delta_{k}\mid^{2}$ is the linear power spectrum of density fluctuations. Comparing with the distribution 
of $\delta_{2}$ at a randomly chosen point which is the one implicitly involved in the derivation of the unconditional 
mass function, we find:

\begin{equation}
P(\delta_{2})=\frac{\exp \left(-\frac{1}{2}\frac{\delta_{2}^{2}}{\sigma_{2}^{2}} \right)}{\sqrt{2\pi}\sigma_{2}}  \nonumber
\end{equation}

We see that, at least for the one point statistics, the field $\delta_{2}$ at a constrained point behaves like
an unconstrained field but with a mean value proportional to $\delta_{l}$ and a modified power spectra. It may
be shown (Rubi\~no et al. 2006) that the joint probability distribution for the field $\delta_{2}$
at several neighboring points follows very closely a Gaussian multivariate with the same mean and power spectra as
the one point distribution.

It is then easy to realize that the conditional mass distribution could be obtained 
through the following substitution in eq.(\ref{eq:eq10}) and a renormalization (see \cite{PatBeta2}):

\begin{equation}
n_{c}(m,\delta_{l},q,Q) \propto n_{uST}(m,\delta'_{c},\sigma'(m))  \label{eq:eq9980}
\end{equation}

\begin{equation}
\delta'_{c} = \delta_{c}-\delta_{l}\frac{\sigma_{12}(m,q)}{\sigma_{1}^{2}}  \nonumber
\end{equation}

\begin{equation}
 \sigma'(m) = \big(\sigma_{2}^{2}(m)-\frac{\sigma_{12}^{2}(m,q)}{\sigma_{1}^{2}}\big)^{1/2}  \nonumber
\end{equation}
$\sigma_{2}$, $\sigma_{12}$ depends on mass through the mass-scale relationship:

\begin{equation}
Q_{2}(m)=\bigg(\frac{m}{3.51\times 10^{11}\Msunh}\bigg)^{1/3}\mpch    \label{eq:eq111}
\end{equation}

\begin{equation}
\sigma_{2}(m)\equiv\sigma(Q_{2}(m))  \nonumber
\end{equation}

\begin{equation}
\sigma_{12}(q,Q,m)\equiv\sigma_{12}(q,Q,Q_{2}(m))  \nonumber
\end{equation}
after this substitution, we obtain the local Lagrangian mass function, $n_{c}(m,\delta_{l},q,Q)$, 
which as we advanced, in practice, is only a function of $q/Q$.

Note that this extending procedure is not compromised with any particular fit to the UMF. Actually, we could
use for example the fit proposed by Jenkins et al. (2001). For our purposes, however, expressions (12) is to be
preferred, because it is very accurate in the mass range we are interested in.

It must be noted that for our expression for the local number density of collapsed objects of mass $m$ to be valid 
this quantity must change little within a distance of the order of the scale corresponding to $m$. 
However, since the scale of variation of the density (for any $m$) is on the order of the void radius, 
it is clear that this condition holds provided that $Q_{2}(m)/Q << 1$.

To carry out the computation we first obtain a fit to $\sigma_{12}(m,q)/\sigma^{2}(Q)$ and find that
it has the form:

\begin{equation}
\frac{\sigma_{12}(q,Q,m)}{\sigma^{2}(Q)}=c_{1}~e^{-c_{2}(\frac{q}{Q})^{2}}   \label{eq:eq110}
\end{equation}
where $c_{1}$ and $c_{2}$ are certain coefficients almost independent of mass for $Q_{2}(m)<<Q$, and
only mildly depending on $Q$. For $Q$ between $3.3$ and $6.6 \mpch$, which include all voids we will 
consider, $c_{2}$ goes from 0.481 to 0.554. So, we may use the following values for $c_{1}, c_{2}$ over 
the whole range:
 
\begin{equation}
c_{1}=1.3212\quad ;\quad c_{2}=0.525    \nonumber
\end{equation}
inserting this in expression \ref{eq:eq9980}  and using \ref{eq:eq999} we find for $\delta_{ns}(\delta_{l},m)$:

\begin{equation}
1+\delta_{ns}(\delta_{l},m)=A(m)e^{-b(m)\delta_{l}^{2}}   \label{eq:eq11} 
\end{equation}

\begin{equation}
A(m)\simeq 1\quad ;\quad b(m)\simeq b'(m)/2
\end{equation}
which provides a very good fit for $\delta_{l} < 0$. $A(m)$ and $b(m)$ are coefficients 
depending only on mass (for values of Q in the relevant range). Note that we may need 
different values of $c_{1}, c_{2}$ for voids substantially larger than those considered 
here. So, the coefficients $A$, $b$ will, in general, depend on $Q$ as well as on $m$. 
The values of $b'(m)$ corresponding to (37) are given in Appendix A.

\section{Mass density and Halo number density profiles in voids}
\label{ref:Sec3}

In computing the void number densities we have used, among other things, our CMF and the spherical collapse.  Here we 
shall have the opportunity of checking separately how these assumptions works in explaining the structure of 
individual voids.

We start with the spherically averaged mass density profile. A given void, 
characterized by its radius, $R$, (i.e. that of the largest sphere it can accommodate) 
and the mean fractional mass density fluctuation within it, $\delta_{0}$, has a unique profile. 
However, over the ensemble of all voids characterized by these two parameters the density profile varies. 
What we want to obtain is the mean profile over this ensemble and its dispersion both parameterized 
by $\delta_{0}$ and $R$.

For the rare voids that we shall consider the void density profile is equal to that for a randomly 
chosen sphere of radius $R$ with inner fractional fluctuation $\delta_{0}$. 
This is so because the fact of whether or not the sphere contains objects 
(of the type defining the void) can modify the properties of the profile only through the 
value of $\delta_{0}$, which we hold by construction fixed.
So, the mean profile within a randomly chosen sphere (with fractional underdensity $\delta_{0}$) 
is practically unbiased with respect to that for a void with the same radius and inner underdensity. 
With this in mind, we may obtain the profile by means of the probability distribution for $\delta(r)$ 
(the mean fractional enclosed density fluctuation) at a distance $r$ from the center 
of a randomly chosen sphere with the condition that $\delta(R)=\delta_{0}$. 
Transforming to the initial conditions with the relationship of the spherical expansion model  
$DL(\delta)$ (eq.(\ref{eq:eq6b})) between the actual
fluctuation, $\delta$, and its linear value, $\delta_{l}$, our problem is reduced 
to obtaining the probability distribution, in the initial field, for the value of 
$\delta_{l}$ at a Lagrangian distance $r(1+\delta)^{1/3}$ (which transforms into present Eulerian 
distance $r$) from the center of a sphere with Lagrangian radius $R(1+\delta_{0})^{1/3}$ and mean inner linear fluctuation 
$\delta_{1}=DL(\delta_{0})$.
Since the initial conditions are assumed to be Gaussian the distribution of $\delta_{l}$ at a fixed value of $q$
conditioned to $DL(\delta(R))=DL(\delta_{0})\equiv\delta_{1}$ is immediately given by:

\begin{equation}
P(\delta_{l}/q,\delta_{1})=\frac{1}{\sqrt{2\pi}} \frac{\exp \left(-\frac{1}{2} \frac{(\delta_{l}-\frac{\sigma_{\rm 
12}(q)}{\sigma_{1}^{2}}\delta_{1})^{2}}{\big(\sigma_{2}^{2}-\frac{\sigma_{12}^{2 }(r)}{\sigma_{1}^{2}}\big)} \right)}{\big(\sigma_{2}^{2}-\frac{\sigma_{12}^{2
}(r)}{\sigma_{1}^{2}}\big)^{1/2}}       \label{eq:eq13}
\end{equation}

\begin{equation}
\sigma_{1}\equiv\sigma(Q) \quad ; \quad Q=R(1+\delta)^{1/3} \nonumber
\end{equation}

\begin{equation}
\sigma_{2}\equiv\sigma(q) \quad ; \quad q=r(1+\delta)^{1/3}  \nonumber
\end{equation}

\begin{eqnarray}
\sigma_{12}(q)~&=&~\frac{1}{2\pi^{2}}\int_{0}^{\infty}\mid\delta_{k}\mid^{2}~W(qk) \nonumber
\\
~&\times&~ W(Qk)~k^{2}~dk  \nonumber 
\end{eqnarray}

Note that this expression gives the conditional probability distribution for
$\delta_{l}$ at a fixed $q$. This is not the conditional probability distribution for the value of 
$\delta_{l}$ corresponding to the value of $\delta$ (through $\delta_{l}=DL(\delta)$), at some fixed $r$, 
which we represent by $\delta(r)$. 
If it were, we could obtain immediately the conditional distribution for $\delta(r)$ 
using expression (41) and the relationship between $\delta$ and $\delta_{l}$. The correct derivation of 
this distribution can be made by a simple (but tedious) argument that we give in Appendix B.
We find for the probability distribution for $\delta$ (at fixed $r$)conditioned to $\delta(R)=\delta_{0}$:

\begin{equation}
P(\delta/r,\delta_{1})= \frac{d}{d\Delta}P_{c}(\Delta)\bigg|_{\Delta=\delta}  \label{eq:eq14}
\end{equation}

\begin{equation}
P_{c}(\Delta)=1-\frac{1}{2}~erfc\bigg[\frac{\mid DL(\Delta)-\frac{\sigma_{12}(q)}{\sigma_{1}^{2}}\delta_{1}\mid}
{\sqrt{2}~(\sigma_{2}^{2}-\frac{\sigma_{12}^{2}(q)}{\sigma_{1}^{2}})^{1/2}}\bigg]   \nonumber
\end{equation}

\begin{equation}
\delta_{1}=DL(\delta_{0}) \quad ; \quad q=r(1+\Delta)^{1/3}    \nonumber
\end{equation}
with it we have for the mean profile, $\bar{\delta}(r)$:

\begin{equation}
\bar{\delta}(r)=\int_{-1}^{\infty}\delta~P(\delta/r,\delta_{1})~d\delta = \int_{-1}^{\infty} (1-P_{c}(\Delta)) ~d\Delta - 
1  \label{eq:eq15}
\end{equation}

This integral must extend only to $\Delta$ values such that the integrand increases monotonically with $\Delta$.
Calling $u(\Delta)$ the argument of $erfc$ in eq.(44) and eq.(\ref{eq:eq15}) one may
check that the solution to equation:

\begin{equation}
u(\Delta)=u    \nonumber
\end{equation}
may have more than one branch. A necessary and usually sufficient condition for equation (\ref{eq:eq14})
to be valid (see Appendix B) is that a branch, $\Delta^{+}(u)$, monotonically increasing with $u$ does exist. 
Other branches correspond to profiles that have experienced a large amount of shell-crossing, so that
expression (\ref{eq:eq14}) is not valid. In order to account only for the relevant $\Delta^{+}(u)$ branch
the above condition must be imposed upon integral (\ref{eq:eq15}).

In an alternative procedure we may lift the mentioned condition on eq.(\ref{eq:eq15}) (the first equation) using for 
$P(\delta/r,\delta_{1})$ the absolute
value of expression (\ref{eq:eq14}) and dividing it into
the integral over $\delta$ of the absolute value of expression (\ref{eq:eq14}), which is larger than $1$ when
there are additional branches. The difference between this procedure and the former gives a clue as to their
accuracy. They are exact only when they agree; otherwise none of them is exact, the former giving a somewhat 
better result. In a similar way we may obtain $\bar{\delta}^{2}(r)$, and the profile dispersion $\sigma_{\delta}(r)$: 

\begin{equation}
\sigma_{\delta}(r) \equiv (\bar{\delta^{2}}(r)-\bar{\delta}^{2}(r))^{1/2}   \nonumber
\end{equation}

As long as the dispersion of the profile is small, which according to the simulations of \citet{Stefan} holds up 
to $15 \mpch$ for the $10 \mpch$ void, the mean actual profile should not differ much from the transformed 
mean linear profile, which we call the most probable profile (in fact, it is very nearly so). 
Now, from eq.(\ref{eq:eq14}) and (44) we see that the most probable profile is essentially given by the center 
of the Gaussian (in eq.\ref{eq:eq13}; i.e. the mean inner profile):

\begin{equation}
\delta_{l}(q)=\frac{\sigma_{12}(q)}{\sigma^{2}(Q)}\delta_{1}  \nonumber
\end{equation}
where $q$ and $Q$ are the Lagrangian radius.
Transforming $\delta$ into $\delta_{l}$ through the spherical model relationship 
$DL(\delta)$ (equation (\ref{eq:eq6})), and using:

\begin{equation}
\frac{\sigma_{12}(q)}{\sigma^{2}(Q)}\simeq e^{-c(\big(\frac{q}{Q}\big)^{2}-1)}\equiv S(q/Q)  \nonumber
\end{equation}
with
\begin{equation}
c=-\frac{1}{4}\frac{dLn\sigma^{2}(Q)}{dLnQ}
\end{equation}
we may write:

\begin{equation}
DL(\delta(r))=\delta_{1}~S(r(1+\delta(r))^{1/3}/Q)    \label{eq:eq16}
\end{equation}
since

\begin{equation}
q(r)=r(1+\delta(r))^{1/3}  \nonumber
\end{equation}


This equation defines implicitly the "most probable" profile $\delta(r,R,\delta_{0})$ parameterized by $R$ and 
$\delta_{0}$. This profile is simply the initial mean profile transformed according with the spherical model. So,
although presented in a somewhat different way, the computation is the same as those found in the literature 
(see e.g. Van de Weygaert \& van Kampen 1993).

It follows from eq.(\ref{eq:eq14}) that through the following substitution in eq.(\ref{eq:eq16}):

\begin{equation}
S(q/Q)\delta_{1} \longrightarrow S(q/Q)\delta_{1} \pm (\sigma^{2}(q)-S^{2}(q/Q)\sigma^{2}(Q))^{1/2}  \nonumber
\end{equation}  
we can obtain the equations for the upper (+) and lower (-) 68 \% confidence level profiles.

The halo number density profiles may now be obtained by combining the mass profiles 
with equation (\ref{eq:eq11}), which gives the fractional fluctuation, $\delta_{ns}(\delta_{l})$, 
of the halo number density, previously to mass motion (i.e. due to the statistical 
clustering of the protohalos) as a function of the mean value of $\delta_{l}$ within the sphere. 
The entire fractional fluctuation, $\delta_{N}$, is given by:

\begin{equation}
1+\delta_{N}(r)=(1+\delta(r))(1+\delta_{ns}(\delta_{l}(r)))   \label{eq:eq17}
\end{equation}

In the approximation leading to equation (\ref{eq:eq16}) (i.e. where we simply transform 
the mean linear profile) the derivation of $\delta_{N}(r)$ (most probable value) 
is particularly simple since, in this case, for a given $r$ there is not only unique $\delta_{N}$ and $\delta$ 
but also unique $\delta_{l}(r)$.
We may then write in eq.(\ref{eq:eq17}) the $\delta$ value given by eq.(\ref{eq:eq16})
and use $DL(\delta(r))$ (expression 11) for $\delta_{l}(r)$, that is:

\begin{equation}
\delta_{N}(r)=(1+\delta(r))A(m)e^{-b(m)(DL(\delta(r)))^{2}}-1  \label{eq:eq18} 
\end{equation}
$\delta(r)$ being the solution to equation (\ref{eq:eq16}) for given values of 
$r$, $R$ and $\delta_{0}$. This gives the most probable halo number density profile parameterized 
by $R$ and $\delta_{0}$. It must be noted that the full probability distribution for $\delta_{N}$ at
a distance $r$ from the center of the void may be obtained through an argument similar to that leading to
eq.(\ref{eq:eq14}). We find:

\begin{eqnarray}
P(\delta_{N}/r,\delta(R)=\delta_{0})=   \nonumber \\
\frac{1}{2}\frac{d}{d\Delta_{N}}erfc\bigg[\frac{\mid
DL(\Delta_{N})-\frac{\sigma_{12}(q)}{\sigma_{1}^{2}}\delta_{1}\mid}{\sqrt{2}~(\sigma_{2}^{2}-\frac{\sigma_{12}^{2}(q)}{\sigma_{1}^{2}})^{1/2}}\bigg]\bigg|_{\Delta_{N}=\delta_{N}}
 \nonumber
\end{eqnarray}
\begin{equation}
q=r(1+D(DL(\Delta_{N})))^{1/3}
\end{equation}
where $\delta_{l}=DL(\Delta_{N})$ is the solution for $\delta_{l}$ as a function of $\Delta_{N}$ of the equation:

\begin{equation}
1+\Delta_{N}=(1+D(\delta_{l}))(1+\delta_{ns}(\delta_{l}))  \nonumber
\end{equation}
with $D(\delta_{l})$ given by eq.(\ref{eq:eq4}) and $\delta_{ns}(\delta_{l})$ given by eq.(\ref{eq:eq11}).
The following relationship:

\begin{equation}
\delta_{N}'(r)=\frac{1}{3}\frac{1}{r^{2}}\frac{d}{dr}r^{3}\delta_{N}(r) \label{eq:eq188}
\end{equation}
between the local fractional fluctuation at $r$, $\delta_{N}'(r)$, 
and the average enclosed fluctuation within $r$, $\delta_{N}(r)$ (also valid between $\delta'(r)$ 
and $\delta(r)$) may be used to obtain the profile of local halo number density. 

\section{Results}
\label{ref:Sec4}

\subsection{Void counting Statistics}

In this subsection, we apply our formalism to compute the mean number densities of voids for several cases,
and compare them with those found by \citet{Stefan} by means of numerical simulations.
In order to make a direct comparison we have applied the formalism to the cases treated in the mentioned work.

They carried out high resolution $N$-Body simulations using the Adaptive Refinement Tree code (ART) of a cube with $80 
\mpch$ side. The total number of particles is $1024^{3}$ which leads to a maximum resolution
of $4 \times 10^{7} \Msunh$ per particle, and a minimum halo mass of $10^{9} \Msunh$; the cosmological parameters
are $\Omega_{m}=0.3$ and $\Omega_{\lambda}=0.7$.

They identified voids following a criteria similar to ours; they considered 
voids as maximal empty spheres in the distribution of dark matter halos (considered as point-like objects).
In that simulations, they found that for voids defined by halos with mass larger than $5 \times 10^{11} \Msunh$, the 
20 largest voids have radii larger than $7.49 \mpch$, the 10 largest have radii larger than $9.2 \mpch$, the 5 
largest, $11.0 \mpch$ and the 3 largest, $11.3 \mpch$.

On the other hand, when the voids were defined by halos with mass larger than $10^{12} \Msunh$, the 20 largest voids 
have radii larger than $6.95 \mpch$, the 10 largest have radii larger than $8.81 \mpch$, the 5 largest, $11.95 \mpch$ 
and the 3 largest, $12.63 \mpch$.
The halo number densities ($\bar{n}$) were $ 7.44\times 10^{-3}(\mpch)^{-3}$ and $4.08 \times 10^{-3}(\mpch)^{-3}$ 
respectively.

The expected number of voids with radii larger than $r$ within a box of size $L (= 80 \mpch)$, $N(r,L)$, is given by:

\begin{equation}
N(r,L) = \int_{r}^{\infty} \bar{V}(r') n(r') dr' \simeq \bar{V}(r) \bar{P}_{0}(r) \nonumber
\end{equation}

\begin{equation}
\bar{V}(r)=(L - 2r)^{3}  \nonumber
\end{equation}
$\bar{V}(r)$ is the available volume for the voids (for their centers) that, since most voids larger than $r$ are 
only slightly larger than $r$ and expression (15) is a good approximation, the last
result follows.

$\bar{P}_{0}(r)$ is given by expression (\ref{eq:eq1}) with $P_{0}(r)$ given by expression (\ref{eq:eq5}). For 
$(1+\delta_{N})$ we have:

\begin{equation}
(1+\delta_{N})=(1+\delta)(1+\delta_{ns})  \nonumber
\end{equation}
where $\delta_{ns}$ is given by eq. (\ref{eq:eq11}) with $A=1$ and $b(m)=b'(m)/2$. For $b'(m)$ we have used the fit 
given in Appendix A.

In table 1 we summarize our results and compare them with the results found by \citet{Stefan}.

We have also estimated the size of the largest void expected in the simulation box at the 90 and 68 per cent of 
confidence level, $\bar{V}(r_{0})\bar{P}_{0}(r_{0})=0.10$ and $\bar{V}(r_{0})\bar{P}_{0}(r_{0})=0.32$ 
respectively. These results are shown in table 2 along with the largest voids actually found in the simulations. 
  
From these results we may infer that expression (\ref{eq:eq1}) gives good results for values of $k$ over 3.5. 
However, for $N > 7$, regardless of the values of $k$, our results differ substantially from those found in the 
simulations. This is due to the fact that the simulation box is small so that it contains only seven or so 
underdense structures (within which voids are found) rare enough $(|\delta_{l}|/\sigma \geq 3)$ for the spherical collapse to be a good 
approximation. To obtain good predictions for voids such that $N > 7$ we must use expression (13) rather than 
eq.(15) and the full expression (Betancort-Rijo \& Lopez-Corredoira 2002) for $P(\delta/r)$ must be used in
expression (9). However, this will rarely be necessary since the constraints imposed on large-scale structure model by void  statistics comes mainly from rare 
voids. 

\begin{table}

\caption{Voids: our results vs. simulations. $N_{p}$ is the Mean number of voids predicted by our formalism 
within the same volume as the simulations one. 
$N_{sim}$ is Number of voids found by \citet{Stefan}. $P_{0}$ is the VPF and $k$ is the rareness of voids 
(defined in eq.(4)). The larger the value of $k$ the rarer is the corresponding void. We can see here that 
as voids become rarer, the analytical predictions are in better agreement with the results from simulations.}

\label{table1} 
\vspace{0.2 cm}
\begin{tabular}{ccccc}  

\hline  \hline
Radius &  &  &  &  \\ 
($\mpch$) &  $P_0$ &  $N_p$ & $N_{sim}$  & $k$   \\  
\hline  \hline

\noalign{\bigskip}
\noalign{$mass=5 \times 10^{11} \msunh$~~~~~$\bar{n}$=0.00744~~b=0.0797}
\noalign{\smallskip}

\hline
11.3  &  0.00101885   &    3.4    &        3     &     4.880  \\                        
11.0  &  0.00149288   &    4.8    &        5     &     4.621  \\
10.8  &  0.00191639   &    6.0    &        7     &     4.471  \\          
9.4   &  0.00984866   &   24.8    &       10     &     3.456  \\
7.4   &  0.072409     &  119.2    &       20     &     2.227    \\
\hline 

\noalign{\bigskip}
\noalign{$mass=1 \times 10^{12} \msunh$~~~~~$\bar{n}$=0.00408~~b=0.1084}
\noalign{\smallskip}

\hline
12.6  &  0.000427927  &    1.8       &     3     &     5.434  \\
11.95 &  0.00107413   &    4.1       &     5     &     4.945   \\
11.2  &  0.00191895   &    7.4       &     7     &     5.035  \\          
8.8   &  0.0423952    &    70.2      &    10     &     2.756   \\
7.0   &  0.193217     &   170.3      &    20     &     1.729   \\

\hline 

\noalign{\bigskip}
\noalign{$mass=2 \times 10^{12} \msunh$~~~~$\bar{n}$=0.002162~~b=0.1389}
\noalign{\smallskip}

\hline
14.05 &  0.00119284   &    2.2      &      3     &     5.432  \\
12.8  &  0.00469144   &    6.9      &      5     &     4.432   \\
11.8  &  0.0125963    &   15.5      &      7     &     3.710  \\
10.8  &  0.0307436    &   30.7      &     10     &     3.050  \\
 7.5  &  0.296388     &  100.5      &     20     &     1.332    \\

\hline

\noalign{\bigskip}
\noalign{$mass=5 \times 10^{12} \msunh$~~~~$\bar{n}$=0.000922~~b=0.3010}
\noalign{\smallskip}

\hline
16.0  &  0.00541489  &     2.6    &     3   &      4.320      \\
14.8  &  0.0136554   &     5.7    &     5   &      3.633      \\
13.5  &  0.0336807   &    11.6    &     7   &      2.957      \\
9.8   &  0.249094    &    35.2    &    10   &      1.394      \\
5.8   &  0.849977    &    19.9    &    20   &      0.400     \\

\hline

\end{tabular}
\end{table}

\begin{table}

\caption{Largest voids. We can see that the agreement between our predictions and 
results from numerical simulations is excellent at $90\%$ Confidence Level.}
\label{table2} 
\vspace{0.2 cm}
\begin{tabular}{cccc}  

\hline  \hline
Mass & $R_{max}$   &  $R_{max}$  &  $R_{max}$    \\
$\msunh$ & 90\% CL &   68\% CL   &  Simulations  \\
\hline  
\noalign{\smallskip} 
$5 \times 10^{11}$ &  13.8   & 13.0   &   13.0     \\
$1 \times 10^{12}$ &  14.2   & 13.55  &   13.97    \\
$2 \times 10^{12}$ &  16.2   & 15.38  &  14.4      \\
$5 \times 10^{12}$ &  20.04   &  18.6   & 20.03    \\

\hline

\end{tabular}
\end{table}

\subsection{Void mass density profiles}

\begin{figure}
\includegraphics[width=\columnwidth]{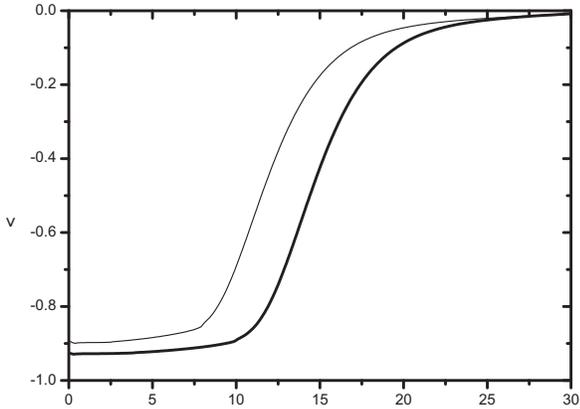}
\caption{The mean enclosed density profiles for voids. The thin line corresponds $R=8 \mpch$ and $\delta_{0}=-0.8667$
while the thick line to $R=10 \mpch$ and $\delta_{0}=-0.9$}
\label{fig:fig2}
\end{figure}

In Fig.(\ref{fig:fig2}) we show the mean density profile using expression (\ref{eq:eq15}) for $R=10 \mpch$ and 
$\delta_{0}=-0.9$, and for $R=8 \mpch$ and $\delta_{0}=-0.8667$  

\begin{figure*}
\includegraphics[width=165mm]{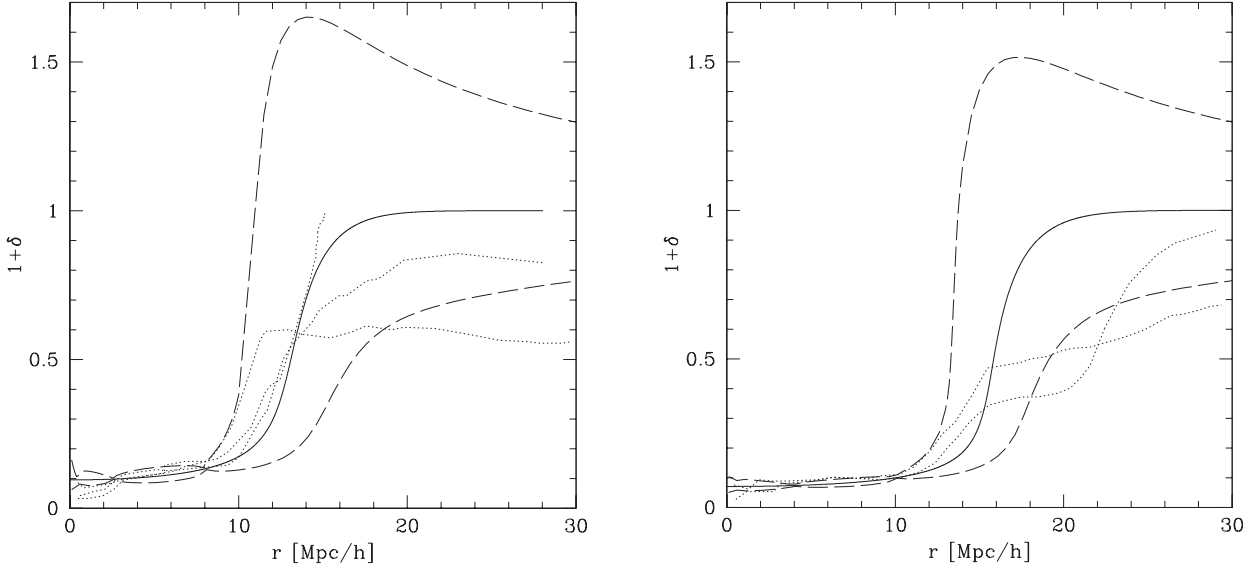}
\caption{The most probable enclosed density profiles for voids. In the left we show the profile for $R=8 \mpch$ and 
$\delta_{0}=-0.86667$ and in the right the plot for $R=10 \mpch$ and $\delta_{0}=-0.9$. The solid line, for both, 
corresponds to the most probable profiles and the dashed line to the 68 \% confidence levels. The dotted lines are 
the profiles found for voids in the numerical simulations.}
\label{fig:fig3}
\end{figure*}

In Fig.(\ref{fig:fig3}) we present the {\emph most probable} profiles for the 
 cases mentioned above including
the 68 \% confidence levels for both profiles, this levels define a quite narrow region up to over $13 \mpch$.

Comparing these two figures it is apparent that, although both profiles are very similar within the voids, the 
mean profile is substantially steeper than the most probable profile at the boundary of the voids. This is 
due to the asymmetry between the upper and lower one sigma profiles.

Note that the profiles given here correspond to an average over all empty spheres with quoted $\delta_{0}$ and $R$, 
while those in the simulations correspond to the largest empty sphere with the same $\delta_{0}$ and $R$. This implies  
that the latter profiles should be somewhat steeper than the former ones for $r$ values slightly larger than $R$ 
(within the sphere, and for $r$ values substantially larger than $R$ they should be equal). It is not difficult to 
account for this effect, but we shall not consider it here since it is not very relevant.

Comparing with simulations (fig. 3 in \citep{Stefan}) we find them to be in very good agreement. 
In particular, for the $R=10 \mpch$ void, the flatness of the profile within the void with a gentle 
descent toward the center ($\delta(0)=-0.93$) is found in our results, as well as the steep rise at the
boundary. This good agreement strongly suggest that the spherical collapse describes correctly 
the dynamics of individual rare voids even when their inner underdensity is quite
low. That the spherical collapse model may give such good results in several cases, 
like the present one, where the degree of spherical symmetry is not that high and 
the tidal field due to the outside matter is not negligible it is an intriguing 
fact that may be explained by the cancellation of the aspherical effects due to the local
matter distribution and the tidal field generated by distant matter (Betancort-Rijo 2004).

It may be checked that for these profiles, both for the most probable one and for the confidence limits, the 
Lagrangian
radius:

\begin{equation}
q(r)=r(1+\delta(r))^{1/3}  \nonumber
\end{equation}
is a monotonically increasing function of $r$. Thus, shell-crossing does not take place and expression
(\ref{eq:eq13}) applies, so that our procedure is consistent. One could think that this implies that, at least for 68 \% 
of the profiles, shell-crossing does not
take place. This is very nearly true, but it must be observed that, in principle, profiles within
the limits may have wiggles, so that shell-crossing could be likely to have taken place; although
even in this case it will not be very relevant, in the sense that eq.(\ref{eq:eq13}) still very nearly
applies, for values of $r$ where the confidence region is narrow.

It must also be noted that it is not strictly true that 68 \% of the profiles are contained within the 
68\% confidence region. This is merely the region generated by the confidence intervals for 
$\delta$ at a fixed value of $r$ as $r$ changes. That is, at a given value of $r$, 68\% of the 
profiles must be within the region although the fraction of profiles that lay within this region 
for all the $r$ values considered may be somewhat smaller. 

\subsection{Halo number density profiles}

\begin{figure}
\includegraphics[width=\columnwidth]{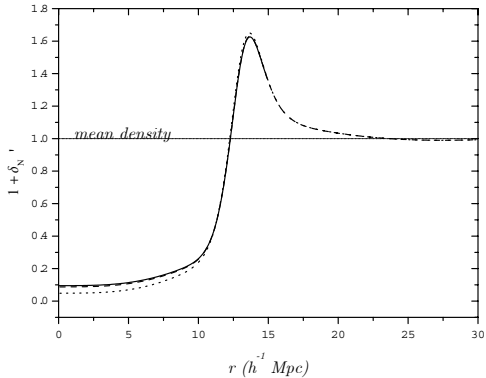}
\caption{In this plot we show the fractional local halo number density within and around a void with $R=8 \mpch$ and 
$\delta_{0}=-0.867$. The dashed line corresponds to halos with mass above $10^{9} \Msunh$ and the dotted line to 
those with mass above $2\times 10^{10} \Msunh$. As a comparison we also show the local mass density profile (full line).}
\label{fig:fig4}
\end{figure}

\begin{figure}
\includegraphics[width=\columnwidth]{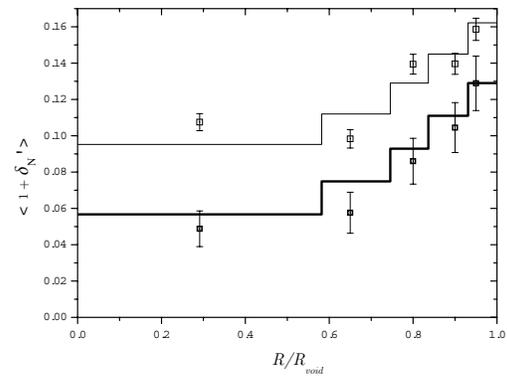}
\caption{The same as Fig. 4, where $<1+\delta_{N}'>$ is the mean halo number density 
within spherical shells with 1/5 of the volume of the void. The thin line correspond to halos with mass above $10^9 
\Msunh$ and the thick line to halos with mass above $2\times10^{10} \Msunh$. $R/R_{void}$ is the distance to the 
center of the void in units of void radius.  The open and half-filled squares correspond to halos with mass above 
$10^9$ and $2\times10^{10}$$\Msunh$ respectively obtained for 5 voids by numerical simulations
(Gottl\"ober et al. 2003). The error bars denote the sampling error.}
\label{fig:fig5}
\end{figure}

We have calculated the most probable local halo number density profile, $\delta_{N}'(r)$, within voids with $R=8 \mpch$ and 
$\delta_{0}=-0.8667$, for masses above $10^{9} \Msunh$ and above $2 \times 10^{10} \Msunh$, which correspond
approximately to halos with circular velocity $20 km/s \leq v_{cir} \leq 55 km/s$ and $55 km/s \leq v_{cir} \leq 120
km/s$ respectively. To obtain $\delta_{N}(r)$ we have used Eq.(56) with $\delta(r)$ given by the most probable mass density profile (Eq.(53))
corresponding to the mentioned values of $R$ and $\delta_{0}$. $\delta_{N}'(r)$ has been obtained from $\delta_{N}(r)$ by means of 
Eq.(61). In Figure (\ref{fig:fig4}) we show the halo number density profiles in and around the void and compared them with the 
local mass density most probable profile (obtained form $\delta(r)$ using expression (\ref{eq:eq188})). It is apparent that $|\delta_{N}|$ is 
larger than $|\delta|$ and the more so the larger the mean mass of the halos in the sample.
In Figure (\ref{fig:fig5}) we show the profiles presented in Fig.(\ref{fig:fig4}), but averaging it over five bins of equal 
volume. Numerical results from \citet{Stefan} are also shown for comparison. 
Although this last results corresponds to a superposition of five different values of $R$ and $\delta_{0}$ the agreement is quite good. 
Note that using our CMF is essential to explain the 
results found in the simulations: the ratio between the local halo number density (for masses above $2\times 10^{10}\Msunh$)
at the center and at the boundary is about 2.6, while for the mean mass density this ratio is 
only 1.54 (for $R=8 \mpch$ and $\delta_{0}=-0.867$). The extra factor 1.69 is due 
to the different statistical clustering of the protohalos at the center and at the 
boundary, that is, to the dependence on position of the local number density of 
protohalos before mass motion (i.e. on Lagrangian coordinates).

\begin{figure*}
\includegraphics[width=165mm]{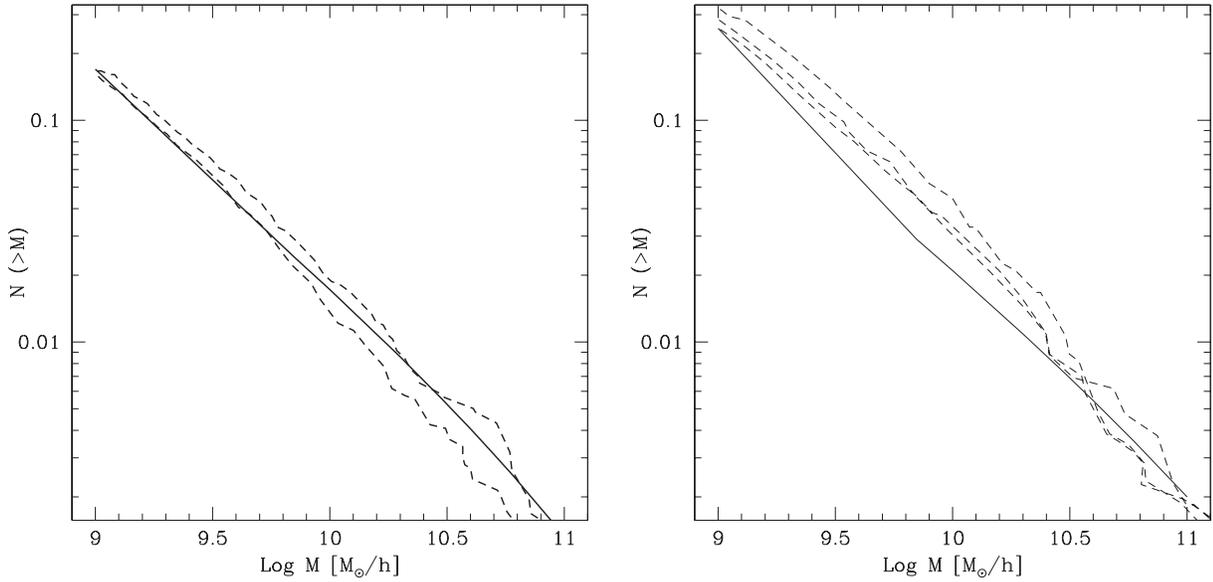}
\caption{Mass functions of halos averaged within two different voids. Left Panel: The
thick full line corresponds to voids with radius $R=10 \mpch$ and mean density,
$\delta_{0}=-0.9$ obtained with our formalism. Also in this panel we show the mass function for 2 voids of the same parameters ($R,
\delta_{0}$) obtained from numerical simulations \citep{Stefan}(thick dashed line). Right Panel: The thin line denote the mass function 
for voids with radius $R=8 \mpch$ and $\delta_{0}=-0.86667$. Again, we show results for 3 voids with the same parameters from the 
numerical simulations. We can see that the agreement of our results with the numerical simulations is very good.}
\label{fig:fig6}
\end{figure*}

In figure (\ref{fig:fig6}) we show the halo mass function for two different voids. Note the excellent
agreement with simulations (fig.(5) in Gottl\"ober et al. 2003).

Summarizing, the distribution of matter and halos of given masses within and around voids 
in simulations may be both reproduced by the combined use of the spherical expansion model 
and our CMF expression. 

\section{Discussion}

So far in this work we have been dealing only with voids defined by dark matter halos. However, the number density of voids 
defined by galaxies may be obtained in the same way as those defined by DM halos. 

To obtain the number density of the latter we implicitly had to determine the relative biasing of halos above certain mass 
with respect to the matter. 
This biasing was responsible for the fact that instead of using in expression (\ref{eq:eq5}) $\delta_{N}=\delta$ 
which correspond to objects distributed like mass, we had to use:

\begin{equation}
1+\delta_{N}=(1+\delta)(1+\delta_{ns}) \label{eq:eq1000}
\end{equation} 
where $1+\delta_{ns}$, which is due to the initial statistical clustering of the protohalos, accounts for, or rather, 
is the origin of the biasing of halos with respect to mass.

$1+\delta_{ns}$ was obtained by studying the dependence of the Lagrangian fractional fluctuation of the number of protohalos 
within a sphere on the linear density fluctuation, $\delta_{l}$, within it.

For voids defined by galaxies of certain type above a given luminosity, $L$, we must use in expression 
(\ref{eq:eq1000}) $1+\delta_{Ls}$, which describes the initial statistical clustering of the protogalaxies, instead of 
$1+\delta_{ns}$. $1+\delta_{Ls}$ is obtained by means of the unconditional, $n_{u}(L)$, and conditional, 
$n_{c}(L,q,Q,\delta_{l})$, luminosity function in the same way as $1+\delta_{ns}$ was derived by means of the 
conditional and unconditional mass functions.

To obtain $n_{u}(L)$ and $n_{c}(L)$, we shall first assume that there exist a universal (independent of the 
environment) Conditional Luminosity Function, $\phi(L/m)$ \citep[CLF]{VanDen}. Then we have:

\begin{equation}
n_{u}(L)=\int_{0}^{\infty}\phi(L/m)~n_{u}(m)~dm \nonumber
\end{equation}	
 
\begin{equation}
n_{c}(L,Q,q,\delta_{l})=\int_{0}^{\infty}\phi(L/m)~n_{c}(m,Q,q,\delta_{l})~dm
\end{equation}	

Integrating over luminosity from $L$ to infinity we obtain $n_{u}(> L)$ and $n_{c}(>L,Q,q,\delta_{l})$. Dividing the 
latter by the former we obtain $1+\delta_{Ls}(q,Q,\delta_{l})$. Averaging over $q$ within the void we finally have:

\begin{equation}
(1+\delta_{Ls}(Q,\delta_{l}))=\frac{3}{Q^{3}}\int_{0}^{Q}\frac{n_{c}(>L,Q,q,\delta_{l})}{n_{u}(>L)}~q^{2}~dq \nonumber
\end{equation}
now, since $n_{c}$ depends on $q$ and $Q$ almost entirely through $q/Q$, in practice, $\delta_{Ls}$ depends only on 
$\delta_{l}$.

So, the number density of voids with radius larger than $r$ defined by galaxies with luminosity larger than $L$  
is given by expression (\ref{eq:eq1}) with $P_{0}(r)$ given by eq. (\ref{eq:eq5}) and $1+\delta_{N}$ given 
by:

\begin{equation}
1+\delta_{N}(\delta_{l})=(1+D(\delta_{l}))(1+\delta_{Ls}(\delta_{l}))  \nonumber
\end{equation} 

More generally, one might consider the plausible possibility that void galaxies are a systematically different
population (Szomoru et al. 1996a; El-Ad \& Piran 2000; Peebles 2001; Rojas et al. 2004) or, in mathematical 
terms, that the Conditional Luminosity Function depends on the environment. Evidence to this dependence on 
theoretical ground have been recently pointed out by Gao et al.(2005) who reported a dependence of halo 
clustering on environment through the environmental dependence on halo formation history. 

Assuming this dependence enters only through the environmental density we should then use $\phi(L/m,\delta_{2})$, 
where $\delta_{2}$ stands for the present linear value of the fractional density fluctuation within a sphere 
centered at the galaxy and with radius $~ 3-4 \mpch$ (this radius should be large enough to define a local environment 
but substantially smaller than the scale length on which these environment change, i.e. large voids).

The only change with respect to the previous case is that now to obtain $n_{u}(L)$ and $n_{c}(L)$ expression (25) must
be multiplied by the probability distribution for $\delta_{2}$ at $q$,
$P(\delta_{2}/q,Q,\delta_{1})$, and integrated over $\delta_{2}$.

Up to here we have been using a non-uniform Poissonian clustering model: the probability per time unit for a galaxy
to form at a given halo may be a function of time and some underlying field, but does not depend on whether or not some
galaxies has previously formed in its neighborhood.
Dark matter halos formed from Gaussian initial conditions may be shown to obey a Poissonian model. This is a consequence of the 
validity for them of the peak-background splitting approximation \citep[BBKS]{BBKS}. However for galaxies this does not need 
to be true if the conditional luminosity function depends on the presence of neighboring galaxies: A general model 
containing these possibilities is given in \citet{Beta00}. To work within this general model we only need to change 
$\exp(-\bar n V)$ in eq.(\ref{eq:eq1}) by $(1+w\bar{n})^{-V/w}$ where $w$ is an additional parameter to be determined from
observations (for Poissonian model $w=0$).

By means of these expressions we may be able to use void statistics to impose constraints on the possible dependence of 
$\phi(L/m)$ on environment, and determine the $w$ value.     

\section{Conclusions}
The extension of the unconditional mass function that we have developed increases 
very substantially the reach of the original useful tool. Not only is the extension 
formally exact under the generally satisfied conditions that we have discussed, but 
it provides the local number density of collapsed objects at any distance from the 
point at which the constraint is evaluated. This is an essential point for the present work as well as for several 
other applications, since the fractional number density varies strongly from the center to the 
boundary of the void (the first value is roughly the cubic root of the second one for any mass). 
Previous procedures for obtaining the conditional mass function assume that the mass 
function and the condition are evaluated at the same point. In principle, one could follow those
procedures to obtain the mass function at any distance from the center of the void. But this is rather
complex and can, at most, provide a good fitting formula. However, as we have 
shown, once the unconditional mass function has been obtained by deriving a fitting formula through the mentioned 
procedures (or any other) and calibrating it by means of numerical simulations, its extension 
to the conditional case can immediately be obtained, without having to repeat the derivation of the 
fitting formula and its calibration.

This formalism has allowed us to obtain the mean fractional number density of 
collapsed objects of given mass within a void as a function of the fractional 
mass density within it. This has been used within the general formalism that 
we have described here to obtain the number density of voids with radius above a given value. 
We have compared our results with those found in numerical 
simulations, checking that for sufficiently rare voids our procedure gives very good results.
Furthermore, using $P(\delta/r)$ as given by the spherical expansion/collapse approximation seems to 
be enough within present uncertainties. Note, however, that we have not checked separately to a
sufficient extent the accuracy of the relationship between $P_{0}(r)$ and $\bar{P}_{0}(r)$ (Eq.(1)) on the one hand and 
the accuracy of our computation of $P_{0}(r)$ on the other. We intend to do this by means 
of more detailed simulations that will allow us to eliminate some minor uncertainties 
thereby increasing the accuracy of our procedure.

One relevant issue that we have not addressed here is the redshift distortions. Note that 
our formalism correspond to real space, while observations are made in redshift space. Recently, 
a procedure for correcting the observed statistics for redshift distortions has been developed (Patiri et al. 2005). 
However, in a future work we intend to complement our formalism so that we could make predictions of voids 
statistics directly in redshift space.

Our formalism provides a simple relation between properties of the galaxy formation 
process and galaxy distribution statistics. Using this relation we may infer those properties 
from the observed distribution. In this manner, in order to asses the consistency of a model
of galaxy formation  with void statistics it is not 
necessary to carry out complex numerical simulations with a large dynamical range 
but only to demand those models to show the properties (i.e. the conditional Luminosity function) 
inferred through our procedure from void statistics. 
Furthermore, the effect of the change of any parameter of the model may be estimated immediately.

\section*{acknowledgments}

We would like to thank the referee for comments and suggestions which 
greatly improved the previous version of the paper.
We thank Stefan Gottl\"ober for productive and useful discussions, 
and also for kindly providing us in electronic
form the statistics of voids from simulations included in Gottl\"ober et al. (2003).
We also thank to Conrado Carretero and Robert Juncosa for helpful 
technical assistance.

\appendix
\section{ Alternative derivation of $\delta_{\lowercase{ns}}$}

In an alternative and more explicit procedure, instead of fitting directly 
the dependence of $\delta_{ns}$ on $m$ and $\delta_{l}$, we first obtain 
the dependence of the fractional halo number density at $q=0$, which we call $\delta_{ns}'$,
on $m$ and $\delta_{l}$. That is, before dealing with the mean fractional 
fluctuation within the sphere of radius $Q$ we deal with its value at the 
center of this sphere. 
We find that $\delta_{ns}'$ may be approximated very accurately (for $\delta_{l}\leq 0$) by:

\begin{equation}
1+\delta_{ns}'(\delta_{l},m)=A'(m)e^{-b'(m)\delta_{l}^{2}}   \label{eq:eq12}
\end{equation}

where for $m$ between $10^{9} \Msunh$ and $2 \times 10^{12} \Msunh$:

\begin{equation}
  A'(m) \simeq 1  \nonumber
\end{equation}
\begin{equation}
  b'(m)=0.0205 + 0.1155 ~\bigg(\frac{m}{3.51 \times 10^{11} \Msunh}\bigg)^{0.5}  \nonumber
\end{equation}

for larger masses we have:

\begin{equation}
  A'(m) \simeq 1  \nonumber
\end{equation}
\begin{equation}
  b'(m)=0.1917 + 0.0198 ~\bigg(\frac{m}{3.51 \times 10^{11} \Msunh}\bigg)  \nonumber
\end{equation}


To obtain the fractional density fluctuation of the number density of collapsed objects 
with masses above $m$, at a distance $q$ from the center of the sphere under consideration, 
$\delta_{ns}(m,\delta_{l},q)$ we simply need to note that:

\begin{eqnarray}
1+\delta_{ns}(m,\delta_{l},q) &=& (1+\delta_{ns}'(m,\frac{\sigma_{12}(m,q)}{\sigma_{12}(m,0)}\delta_{l})) \\ 
&\times& \bigg(\frac{2+\frac{\sigma^{2}_{1}}{\sigma^{2}_{2}}(\frac{\sigma_{12}(m,q)}{\sigma_{1}})^{2}} 
{2+\frac{\sigma^{2}_{1}}{\sigma^{2}_{2}}(\frac{\sigma_{12}(m,0)}{\sigma_{1}})^{2}}\bigg) \nonumber
\end{eqnarray}

where in the first parenthesis we have noted that in the procedure for extending the UMF expression the dependence on 
$\delta_{l}$ enters essentially through 
 
\begin{equation}
\frac{\sigma_{12}(m,q)}{\sigma^{2}(Q)}\delta_{l}  \nonumber
\end{equation}

which as we have seen (eq.\ref{eq:eq110}) is independent of $m$.
The second parenthesis accounts for the differences between the expressions that 
substitute $d\ln \sigma/dm$ in the extensions of UMF expression at $q=0$ and at a given $q$.
This parenthesis is, for the values we shall use, very close to $1$ and may be neglected.
Having $\delta_{ns}(m,\delta_{l},q)$ at any $q$ we may immediately obtain $\delta_{ns}(m,\delta_{l})$ 
by computing its mean value within the sphere:

\begin{equation} 
1+\delta_{ns}(m,\delta_{l})=3~\int_{0}^{1}~A'(m)e^{-b'(m)\delta_{l}^{2}e^{-1.05u^{2}}}u^{2}du  \nonumber
\end{equation}

here, we have used eq.(\ref{eq:eq999}) and eq.(A1), setting $u \equiv q/Q$.
We find again that this expression may be accurately fitted by eq.(\ref{eq:eq12}) with $A(m)\simeq A'(m)\simeq 1$ and 
$b(m)$ approximately equal to $b'(m)/2$ in all the range of masses considered. 
$\delta_{ns}$ is the mean (within $Q$) fractional fluctuation of the number density in 
Lagrangian coordinates (i.e., before the halos move along with mass). 
This fluctuation is due to the statistical clustering, in the initial field, of the protohalos under
consideration (those with mass above $m$). 

\section{ Derivation of $P(\delta/r,\delta_{1})$}

To obtain expression (\ref{eq:eq14}) for the probability distribution for 
the mean value of $\delta$ within $r$, we compute first the probability that $\delta\leq\Delta$ (where $\Delta$ is 
some given value), which we represent by $P_{c}(\Delta)$. If at a certain $q$ value, $q_{0}$, $\delta_{l}(q_{0})$ 
(the linear value of the fractional density fluctuation within Lagrangian radius $q$) were equal 
to $DL(\Delta)$ (the linear value corresponding through spherical collapse 
to an actual value $\Delta$), then the sphere concentric with 
the void and with Lagrangian radius $q_{0}$ show a mean inner fractional fluctuation 
equal to $\Delta$ corresponding to an expansion by a factor $(1+\Delta)^{-1/3}$ and, if its present
radius is $r$, we must have $q_{0}=r(1+\Delta)^{1/3}$. If $\delta_{l}(q_{0})$ were 
less than $DL(\Delta)$ then the actual $\delta$ 
within this sphere (with Lagrangian radius $q_{0}$) will be 
less than $\Delta$. So, the sphere with Lagrangian radius, $q_{0}$, would have expanded 
by a factor larger than $(1+\Delta)^{-1/3}$ its present size being then 
larger than $r$. 

Thus, the sphere with present radius $r$ would have expanded from an initial one 
with $q'<q_{0}$ (neglecting shell crossing) and, assuming that the profile is monotonically increasing (which is 
obviously true for the most probable profile, but is also so for any realization 
with non-negligible probability), $\delta_{l}(q')$ shall be less than $\delta_{l}(q)$, and consequently, 
$\delta(r)$ shall be less than $\Delta$. We may then write: 

\begin{equation}
P_{c}(\Delta)~\equiv~P(\delta\leq\Delta,/r,\delta_{1})~=~P(\delta_{l} \leq DL(\Delta),/q_{0}=r(1+\Delta)^{1/3},\delta 
_{1})  \nonumber
\end{equation}

\begin{equation}
P(\delta/r,\delta_{1})=\frac{d}{d\Delta}P_{c}(\Delta)\bigg|_{\Delta=\delta}  \nonumber
\end{equation}
where (B1) denote the fact that the probability that $\delta$ be smaller than $\Delta$ at $r$
is, by the above arguments, equal to the probability that $\delta_{l}$ be smaller than $DL(\Delta)$ at $q_{0}$.
But for fixed $q_{0}$ and with the condition $\delta(R)=\delta_{0}$, 
$\delta_{l}$ follows distribution (\ref{eq:eq13}), so, expression (43)
follows immediately.

We have assumed above that shell crossing does not occur, however this may be shown 
to be the case in the relevant applications. Note that the relevant shell crossing 
here is that on the scale of the void, that is the shell crossing experienced by 
the mass field filtered on the scale of the void. On much smaller scales there is 
obviously shell crossing, but this is irrelevant to our argument.

\end{document}